\documentclass[useAMS]{mn2e}
\usepackage{amsmath}
\usepackage{textcomp}
\usepackage{amssymb}
\usepackage[pdftex]{graphicx}
\usepackage{amssymb}

\usepackage[margin=.8in, paperwidth=8.5in, paperheight=11in]{geometry}
\usepackage{multicol}
\usepackage{multirow}
\usepackage{epsfig}
\usepackage{graphicx}
\usepackage{dblfloatfix}

\font\tenbg=cmmib10 at 10pt

\def \rvecphi{{\hbox{\tenbg\char'036}}}

\def\lesssim{\mathrel{\hbox{\rlap{\hbox{\lower4pt\hbox{$\sim$}}}\hbox{$<$}}}}
\def\gtrsim{\mathrel{\hbox{\rlap{\hbox{\lower4pt\hbox{$\sim$}}}\hbox{$>$}}}} 
\usepackage{accents}
\newcommand*{\dt}[1]{%
  \accentset{\mbox{\Large\bfseries .}}{#1}}

\begin{document}

\title{Kelvin-Helmholtz Instability of Counter-Rotating  Discs}

\author{Dan Quach \& R.V.E. Lovelace}

\author[Dan Quach, Sergei Dyda, and  Richard V.E. Lovelace]{
Dan Quach$^1$, Sergei Dyda$^1$, and Richard V. E. 
Lovelace$^{2}$\\
$^{1}$Department of Physics, Cornell University, Ithaca, NY 14853, USA\\
$^{2}$Department of Astronomy, Cornell University, Ithaca, NY 14853, USA}

\maketitle

\begin{abstract}

    Observations of galaxies and models of accreting systems point to
the  occurrence of counter-rotating discs where the inner part of
the disc ($r<r_0$)  is co-rotating and the outer part is counter-rotating.
This work analyzes the linear stability of radially separated 
co- and counter-rotating thin discs.   The strong instability found is
the supersonic Kelvin-Helmholtz instability.  The growth rates are of the order of or larger than the angular rotation rate at the interface.   The instability is absent if there is no vertical dependence of the 
perturbation.   That is, the instability is essentially
three-dimensional.   The nonlinear evolution of the instability is
predicted to lead to a mixing of the two components, strong heating of
the mixed gas, and vertical expansion of the gas, and annihilation of the angular momenta of the two components.   As a result the heated gas will free-fall towards the disc's center over the surface of the inner disc.

\end{abstract}


\section{Introduction}

   Commonly considered models of accretion discs have gas
rotating in one direction with a turbulent viscosity acting to
transport angular momentum outward and matter inward.
   However, observations indicate that there are  more complicated, possibly transient,  disc structures on a galactic scale.
   Observations of normal galaxies have revealed co/counter-rotating gas/gas, gas/stars, and stars/stars discs
   in many galaxies of all morphological types ---
ellipticals, spirals, and irregulars (Rubin 1994a, 1994b; Galletta
1996; recent review by Corsini 2014).   There is 
limited evidence  of  counter-rotating gas components in discs around stars (Remijan and Hollis 2006; Tang et al. 2012).
     Theory and simulations of
 co/counter-rotating gas/gas discs around stars and black holes predict
enormously enhanced accretion rates resulting from the
presence of the two components which can give rise to
outbursts from these discs (Lovelace and Chou 1996; Kuznetsov et al.
1999;  Nixon, King, and Price 2012; Dyda et al. 2014).

   Counter-rotation in disc galaxies appears in a variety
of configurations including cases where (1) two cospatial 
populations of stars rotate in opposite directions,
(2) the stars rotate in one direction and the gas rotates
in the opposite direction, and (3) two spatially separated
gas discs rotate in opposite directions (Corsini 2014).
  Of particular interest here is the
 example of case (3), the ``Evil Eye'' galaxy
NGC 4826 in which the direction of rotation of the gas
reverses going from the inner ($180$ km s$^{-1}$) to the outer disc
($-200$ km s$^{-1}$) with an inward radial accretion speed of
more than $100$ km s$^{-1}$in the transition zone, whereas the
stars at all radii rotate in the same direction as the gas in the
inner disc, which has a radius of $\sim 1200$ pc (Braun et al. 1994;
Rubin 1994b).   

     On a stellar mass scale,  Remijan and Hollis (2006)
 and Tang et al. (2012)  found evidence of  accretion of counter-rotating gas in star forming systems. 
    Accreted counter-rotating matter may
encounter an existing co-rotating disc as in low-mass X-ray
binary sources where the accreting, magnetized rotating
neutron stars are observed to jump between states where
they spin-up and those where they spin-down. 
Nelson et al. (1997) and Chakrabarty et al. (1997) have proposed
that the change from spin-up to spin-down results from a
reversal of the angular momentum of the wind-supplied
accreting matter.  The interface
between the oppositely rotating components would be subject
to the Kelvin-Helmholtz instability.

       In the formation of massive black holes in the nuclei
 of active galaxies, King and Pringle (2006, 2007) argue
 that in order for the rapid growth of the massive black holes
 (observed at high red-shifts) to be compatible with the
 Eddington limit on their emission, the gas accretion is from
 a sequence of clouds with randomly oriented angular momenta.
 This process would lead to the formation of a disc with the
 rotation direction alternating as a function of radius.  

    Section 2 develops the theory for the stability analysis and
 obtains explicit growth rates for the supersonic Kelvin-Helmholtz instability  between radially separated co- and counter-rotating thin disc
 components.  Section 3 discusses possible nonlinear 
 consequences of the instability.  Section 4 gives the main
 conclusions.

\section{Theory}

\subsection{Disc Equilibrium}

      We consider the stability of a thin counter-rotating
 disc around a star or black hole.  Figure 1 shows
 the rotation curve we consider.
  We use an inertial cylindrical $(r,\phi,z)$ coordinate system.
      The equilibrium  has
($\partial/\partial t=0$) and ($\partial/\partial \phi = 0$),
with the flow velocity
${\bf u} = u_\phi(r) \hat{\rvecphi~}=r\Omega\hat{\rvecphi~}$.  
   That is, the accretion velocity $u_r$ and the vertical
velocity $u_z$  are assumed negligible compared with $u_\phi$.
The equilibrium flow satisfies $-\rho r \Omega^2=
-dp/dr+\rho g_r$, where $\rho$ is the density, $p$ the pressure,
$g_r= -\partial \Phi/\partial r$ and
$\Phi$ the gravitational potential.
   We consider  thin discs with half-thickness 
   $h \approx (c_s/u_\phi)r \ll r$, where $c_s$ is the sound speed.
In this limit $\Omega =(g_r/r)^{1/2}[1+{\cal O}(h^2/r^2)]$.  We
consider  Keplerian discs where $g_r=GM/r^2$  (with $M$
the mass of the central object) as well as more general
discs, for example, flat rotation curve discs
relevant to galaxies where $g_r \propto  r^{-1}$.
      We neglect the vertical structure of the equilibrium disc
 assuming $\partial/\partial z =0$ for the equilibrium quantities
 (e.g., $\rho$, $p$, and $u_\phi$).
   Our treatment parallels that of Lovelace, Turner, and Romanova 
 (2009; denoted LTR09), but it generalizes that work to include
 the $z-$dependence of the perturbations.

\subsection{Perturbation Equations}

The perturbed quantities are:
the density,
$\tilde{\rho} = \rho +
\delta \rho(r,\phi,z,t)$;
the pressure is
$\tilde{p} = p+\delta p(r,\phi,z,t)$; and
the  flow
velocity is $\tilde{\bf u} =
{\bf u} +\delta {\bf u}(r,\phi,z,t)$ with
${\bf \delta u} =
(\delta u_r,\delta u_\phi, \delta u_z)$.
   We neglect the self-gravity of the disc
as commented on at the end of this
section.
    The equations for the
perturbed flow are
\begin{equation}
{D \tilde{\rho}\over Dt}
+ \tilde{\rho}~ {\bf \nabla}\cdot
\tilde{\bf u} = 0~,
\end{equation}
\begin{equation}
{D \tilde{\bf u}\over Dt}  =
-{1\over \tilde{\rho}}
{\bf \nabla} \tilde{p}
 - {\bf \nabla}\Phi ~,
\end{equation}
\begin{equation} 
{D S\over Dt} = 0~,
\end{equation}
where $D/Dt \equiv \partial /\partial t
+ \tilde{\bf u}\cdot {\bf \nabla}$, and where
 $S \equiv {\tilde p}/(\tilde \rho)^\gamma$
is the entropy of the disc matter.

\begin{figure}
    \centering
    \includegraphics[width=0.3\textwidth]{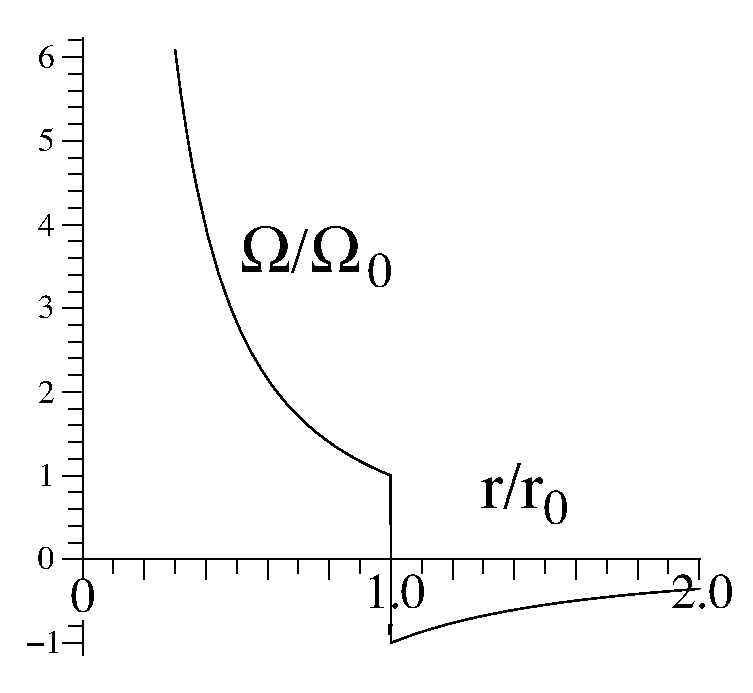}
    \caption{Angular velocity of the counter-rotating Keplerian disc
where $g_r=GM/r^2$ with $M$ the mass of the central object.}
\end{figure} 

We consider perturbations $\delta p$, $\delta \rho$.. of the form 
\begin{equation}
 f(r)\exp
(im\phi+ik_z  - i\omega t),
\end{equation}  
where $m=0,1,2,..$ is the azimuthal
mode number, $\omega$ the angular frequency, and $k_z$ the wavenumber in the $z-$direction.
From equation (1), we find
\begin{equation}
i\Delta \omega~ \delta \rho =
{\bf \nabla} \cdot
(\rho ~\delta {\bf u})~,
\end{equation}
where
$\Delta \omega(r) \equiv \omega - m \Omega(r)$ and
$\Omega = u_\phi/r$.
From equation (2),
\begin{equation} 
i\Delta \omega \delta u_r +
2 \Omega \delta u_\phi =
{1\over \rho}{\partial \delta p\over \partial r}
-{\delta \rho \over \rho^2} {d p
\over dr}~,
\end{equation}
\begin{equation}
i\Delta \omega \delta u_\phi -
{\kappa^2 \over 2 \Omega}
\delta u_r = ik_\phi
 {\delta p \over \rho}~,~~~~~~~~~~~~~~
\end{equation}
\begin{equation}
i\Delta \omega \delta u_z = ik_z
 {\delta p\over \rho}~.~~~~~~~~~~~~~~
\end{equation}
Here, 
\begin{equation}
\kappa \equiv
\left[{1\over r^3} {d(r^4 \Omega^2)\over dr}\right]^{1\over 2}
\nonumber
\end{equation}
 is the radial
epicyclic frequency and $k_\phi \equiv m/r$
is the azimuthal wavenumber.
    From equation (3) we have
\begin{equation}
\delta p = c_s^2 -{i\rho c_s^2 \over \Delta \omega L_S}\delta u_r~,
\end{equation}
where 
\begin{equation}
{1\over L_S} \equiv {1\over \gamma}{d \ln(S) \over dr}~,\quad\quad
c_s\equiv \left({\gamma p\over \rho}\right)^{1/2}~,
\end{equation}
with $L_S$ the length scale of
the disc's  entropy variation, and $c_s$ the sound speed.

   Introducing $\Psi \equiv \delta p/\rho$,  equations (6)-(8) 
can be rewritten as
\begin{equation}
\rho \delta u_r =i{\cal F}
\left[{\Delta \omega \over \Omega}
\left({\partial \Psi \over \partial r} 
-{\Psi \over L_*}\right)-2 k_\phi \Psi \right]~,
\end{equation}
\begin{multline}
\rho \delta u_\phi = \mathcal{F} \Bigg\{\frac{\kappa^2}{2 \Omega^2}
\left( \frac{\partial \Psi}{\partial r} 
-\frac{\Psi}{L_S}\right) \\
-k_\phi\left( \frac{\Delta \omega}{\Omega}
+\frac{d p/d r}{\rho \Omega \Delta \omega L_S}
\right)\Psi \Bigg\}~,
\end{multline}
\begin{equation}
\rho \delta u_z = \rho \frac{k_z}{\Delta \omega} \Psi,
\end{equation}
where
\begin{equation}
{\cal F} \equiv {\rho \Omega \over {\cal D}}~,
\end{equation}
and
\begin{equation}
{\cal D}=\kappa^2- (\Delta \omega)^2 -{dp/dr
\over \rho L_S}~.
\end{equation}
Equations (9) and (11) can be combined to give
\begin{equation}
\delta \rho= {\rho \Psi \over c_s^2}
-{{\cal F} \over \Delta \omega L_S}
\left[{\Delta \omega \over \Omega}
\left({\partial \Psi \over \partial r}-{\Psi \over
L_S}\right)-2k_\phi\Psi\right].
\end{equation}

Equations (11)-(13) and (16) can now be substituted into
equation (5) to give   a single differential equation 
for $\Psi(r)$,
\begin{multline}
\frac{1}{r}\frac{\partial}{\partial r} \left(\frac{r \mathcal{F}}{\Omega} \frac{\partial \Psi}{\partial r}\right) =
\bigg[{\rho \over c_s^2}+  {k_\phi^2{\cal  F} \over \Omega} \\
+{1\over r}{\partial \over \partial r}
\left({r{\cal F} \over \Omega L_S}\right)
+{{\cal F} \over \Omega L_S ^2}
\bigg]\Psi +~\left[2k_\phi{\cal F}{d \ln( g{\cal F}) \over dr}\right]
{\Psi \over \Delta \omega} \\
+~\bigg[{k_\phi^2(dp/dr){\cal F}
\over \rho \Omega L_S} - \rho k_z^2 \bigg]{\Psi \over (\Delta \omega)^2}~,
\end{multline}
where $g\equiv \exp(2\int dr/L_S)$.   This equation generalizes
equation (13) of LTR09 by including the $z-$dependence of
the perturbations which contributes the last term in this 
equation. 

  The condition neglecting the self-gravity  is that the Toomre
parameter $Q=\kappa c_s/(\pi G\Sigma)$
be larger than about $r/h$, where $h$ is the disc's half-thickness
and $\Sigma =2h\rho $ is its surface mass density (Lovelace
\& Hohlfeld 2013).

\subsection{Specific Model}

     In order to simplify the analysis we consider barotropic
disc matter where $L_S^{-1} \rightarrow 0$ so that $g=1$.
  Further, we consider  the counter-rotation curve shown in
Figure 1, namely,
\begin{equation}
   \Omega = \left\{
     \begin{array}{lr}
       (g_r/r )^{1/2} &  r < r_0~, \\
       -(g_r/r)^{1/2} &  r > r_0~.
     \end{array}
   \right.
\end{equation}
   We can then solve equation (17) for $\Psi_-$ for  $r<r_0$ and $\Psi_+$  for $r>r_0$ and then match the solutions to the jump-condition across $r=r_0$ which follows from the equation.

We consider solutions of the form
\begin{equation}
\begin{array}{lr}
\Psi_- ~~\propto ~~\exp\big[+k_{r-}(r-r_0) \big]~, & r<r_0 \\
\Psi_+ ~~\propto ~~\exp\big[-k_{r+}(r-r_0)\big]~, ~~&~~ r>r_0
\end{array}
\end{equation}
We assume $|k_{r\pm}|r_0 \gg1$ so that radial variations
of the equilibrium quantities can be neglected.   
In order that the perturbations decay going away from $r_0$,
we must have $\Re(k_{r\pm}) >0$, which can be checked
a posteriori.   

   For $r \gtrless r_0$, equation (17) gives the dispersion relations
\begin{equation}
c_s^2(-k_{r\pm}^2 + k_\phi^2)=[\kappa^2 - (\Delta \omega_\pm)^2]\left[\frac{k_z^2 c_s^2}{(\Delta \omega_\pm)^2} - 1 \right]~,
\end{equation}
where $\Delta \omega_\pm \equiv \omega-m\Omega_\pm$.
   For $k_z=0$, the dispersion relations take the more
familiar form $(\Delta \omega_\pm)^2 = \kappa^2
+c_s^2 (-k_{r\pm}^2 +k_\phi^2)$.

      It is useful to introduce 
the following dimensionless variables,
\begin{equation}
\overline{k}_z = k_z h, ~~ \overline{k}_\phi = k_\phi h, ~~ \overline{\Omega} = \frac{\Omega}{\Omega_0},~~ 
\overline{\kappa} = \frac{\kappa}{\Omega_0},~~  \overline{\Delta \omega} = \frac{\Delta \omega}{\Omega_0}~,
\end{equation}
where $h=c_s/\Omega_0$ and $\Omega_0 = \sqrt{g_r(r_0)/r_0}$.
   Dropping the overbars gives $k_{r\pm}(\omega)$,
\begin{equation}
k_{r\pm}^2(\omega) = k_\phi^2 - [\kappa^2 - (\Delta \omega_\pm)^2]\left[\frac{k_z^2}{(\Delta \omega_\pm)^2} - 1 \right]~.
\end{equation}

     Multiplying equation (17) by $r$,  integrating over $r$ from
$r_0-\epsilon$ to $r_0+\epsilon$, letting $\epsilon \rightarrow 0$,
and taking into account that $\Psi(r)$ is continuous across $r_0$
gives
\begin{equation}
\frac{k_{r+}}{\mathcal{D}_+} + \frac{k_{r-}}{\mathcal{D}_-} = \mathcal{R}(\omega)~,
\end{equation}
where
\begin{equation}
\mathcal{R}(\omega) = k_\phi \left(\frac{1}{\mathcal{D}_+} + \frac{1}{\mathcal{D}_-}\right)\left(\frac{1}{\Delta \omega_+} + \frac{1}{\Delta \omega_-}\right)~.
\end{equation}
  The left-hand side of equation (23) is from the
left-hand side of equation (17).  The right-hand side,   ${\cal R}$ ,
is from the right-hand term of equation (17) involving $d{\cal F}/dr$
which is $d{\cal F}/dr = ({\cal F}_+ -{\cal F}_-)\delta(r-r_0)$.

       For the assumed thin  discs,  $\overline{\Omega} = 1+{\cal O}(h^2/r_0^2)$ for  Keplerian discs and other rotation curves.
       On the other hand,
 \begin{equation}
\mathcal{D}_\pm =\left\{
\begin{array}{lr}
 1 -(\Delta \omega_\pm)^2 ~~~~{\rm Keplerian~,}\\
  {\kappa}^2 -(\Delta \omega_\pm)^2 ~~{\rm other ~rotation ~curves~,}
  \end{array}
  \right.
\end{equation} 
and $\Delta \omega_\pm =\omega \pm m$ with $\omega$ measured in units of $\Omega_0$.

   Equation (23) can be squared twice to obtain
\begin{equation}
G(\omega) =\left(\frac{k_{r+}^2}{\mathcal{D}_+^2} - \frac{k_{r-}^2}{\mathcal{D}_-^2} - \mathcal{R}^2\right)^2 - \frac{4 k_{r-}^2 \mathcal{R}^2}{\mathcal{D}_-^2} = 0~.
\end{equation}
   The possibility of spurious roots of this equation not
satisfying equation (23) has to be checked a posteriori.
Equation (26) can be expanded out as a $16$th order polynomial
in $\omega$.

     Important properties of $G(\omega)$ are readily verified:  
Firstly, for $m=0$ the perturbation is stable,  ${\cal R}=0$ and therefore $k_{r+}=k_{r-} =0$ so that either $\omega=\pm \kappa$ (radial epicyclic motion) or $\omega = \pm k_z$
(vertically propagating sound wave).    For this reason we consider $m\geq 1$ in the following.
   Secondly,  $G(\omega) = G(-\omega)$.   Thirdly,
$[G(\omega)]^* = G(\omega^*)$.  It follows that the solutions for $\omega$ of
$G(\omega)=0=[G(\omega)]^*$ are either purely real or purely imaginary.

    The roots of $G(\omega)$ can be located by making a
contour plot of
\begin{equation}
F(\omega)=\big\{\Re[G(\omega)]\big\}^2 
+ \big\{\Im[G(\omega)]\big\}^2~,
\end{equation}
in the $\omega=\omega_R + i\omega_I$ plane with Maple or
Mathematica. Figure 2 shows the the unstable roots for low
$m$ values for a Keplerian disc with  $h/r_0 =0.1$.   There is stability 
($\omega=$ real) for $k_z < k_{zc}(m)$.
For this range of  $k_z$, the $\omega$ roots are real.   
    We have verified that for $k_z>k_{zc}$  the roots satisfy equation
(23) as well as the conditions $\Re(k_{r\pm}) >0$ needed for the modes to be  localized around $r_0$.  Figure 3 shows the real and imaginary
parts of $k_{r\pm}$ as a function of $k_z$ for $m=1$ also
for a Keplerian disc with $h/r_0=0.1$.

\begin{figure}
    \centering
    \includegraphics[width=0.4\textwidth]{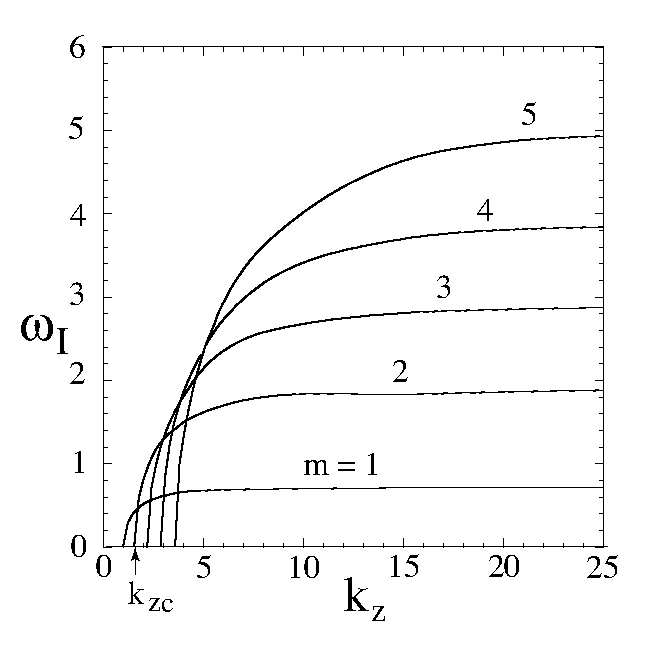}
    \caption{Wavenumber dependence of the growth rate of the instability, $\Im(\omega)=\omega_I$, for different values of $m$ for a Keplerian disc with
$h/r_0 =0.1$.  Note that $\omega_I$ is in units of $\Omega_0$ and $k_z$ is in units of $h^{-1}$.}
\end{figure} 

\begin{figure}[b]
    \centering
    \includegraphics[width=0.37\textwidth]{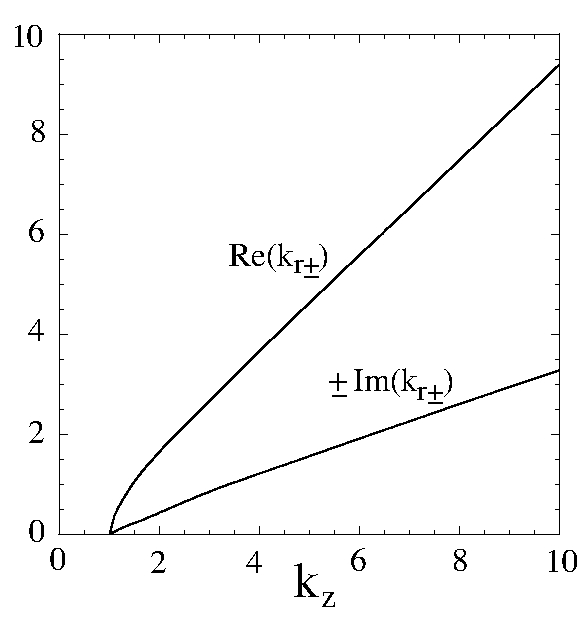}
    \caption{Radial wavenumber, $\Re(k_{r\pm})$ and $\Im(k_{r\pm})$, as a function of $k_z$ for $m = 1$ for a Keplerian disc with$h/r_0=0.1$.    The wavenumbers are in units of $h^{-1}$.
    }
\end{figure} 

     A significant simplification of  the dispersion relation (26) is possible in the limit where $k_z \gg k_\phi = m(h/r_0)$.   In this limit,
\begin{equation}
k_{r\pm} = - {\cal D}_\pm \left(\frac{k_z^2}{\Delta \omega_\pm^2} - 1 \right)~,
\end{equation}
and
\begin{equation}
\frac{k_{r+}^2}{\mathcal{D}_+^2} = \frac{k_{r-}^2}{\mathcal{D}_-^2}~.
\end{equation}
By combining equations (28) and (29) we obtain a function with roots that can be determined analytically.   By rewriting this function as an ordinary polynomial, we obtain a $9$-th order polynomial.
   Factoring out the $\omega = 0$ root gives the $8$-th order polynomial
\begin{multline}
\omega^8 - (2{\kappa}^2 + 2 k_z^2 + 4 m^2)\omega^6 + ({\kappa}^4 + 5 k_z^2 {\kappa}^2 \\
+ 2 m^2 {\kappa}^2 + 2 k_z^2 m^2 + 6 m^4)\omega^4 + (-4 k_z^2 {\kappa}^4- 2 m^2 {\kappa}^4+ 6 k_z^2 m^2 {\kappa}^2 \\
+ 2 m^4 {\kappa}^2 + 2 k_z^2 m^4 - 4 m^6)\omega^2 + (k_z^2 {\kappa}^6 - 4 k_z^2 m^2 {\kappa}^4+ m^4 {\kappa}^4 \\
+ 5 k_z^2 m^4 {\kappa}^2- 2 m^6 {\kappa}^2 - 2 k_z^2 m^6 + m^8) = 0~.
\end{multline}
Rewriting the equation in this form removes the duplicate solutions from the original polynomial.   There are four real roots, $\omega_R=\pm(m\pm \kappa)$ which are the normal modes of a unidirectional  disc, and the four roots
\begin{equation}
\begin{array}{lr}
\omega = \pm\left[k_z^2 + m^2 + \sqrt{k_z^2 (-{\kappa}^2 + k_z^2 + 4 m^2)}\right]^{1/2}~,\\
\omega = \pm\left[k_z^2 + m^2 - \sqrt{k_z^2 (-{\kappa}^2 + k_z^2 + 4 m^2)}\right]^{1/2}~.\\
\end{array}
\end{equation}
   The first set of roots give stable oscillations.  The second set 
 gives purely imaginary $\omega$ values.  The unstable root
 is
\begin{equation}
\omega_I = \left[{\sqrt{k_z^2(k_z^2+4m^2-{\kappa}^2)}-m^2-k_z^2}\right]^{1/2}~.
\end{equation}
For $k_z$ large compared with unity,  $\omega_I \rightarrow m$.

Solving for $\omega_I = 0$ gives the threshold value 
\begin{equation}
k_{zc} = \Big[\frac{m^4}{2m^2 - {\kappa}^2}\Big]^{1/2}~.
\end{equation}
This expression agrees accurately with the solutions obtained
from the full dispersion relation (equation 23).
    Note that for a Keplerian disc, $\kappa =1$, so that all of the
 modes $m=1,2,..$ are unstable.   For flat rotation curves relevant
 to disc galaxies, $u_\phi =$const, so that $\kappa =\sqrt{2}$
 and consequently the mode $m=1$ is stable.   This mode is
 also stable for a rigid rotation curve  disc where $\kappa =2$.
    Note that the dependence of the growth rate on $\kappa$
 shows the dependence of the instability on the curvature
 of the disc.

     The growth rates found here are in approximate agreement with 
 the predictions for a planar vortex sheet  $\omega^2 =
( k_\| c_s)^2[1+{\cal M}_p^2/4 -(1+{\cal M}_p^2)^{1/2}]$, where 
${\cal M}_p \equiv (k_\phi/k_\|){\cal M}_T \leq {\cal M}_T$ 
with $k_\| = \sqrt{k_\phi^2+k_z^2}$
 the wavenumber parallel to the interface, 
and ${\cal M}_T =2 r_0\Omega_0/c_s$ the Mach number of the total velocity jump (see, e.g., Choudhury \& Lovelace 1984).   
    In this case, instability $\omega^2 <0$ occurs for ${\cal M}_p < 2^{3/2}$ which corresponds to $k_z h > m(1/2 -h^2/r_0^2)^{1/2} \approx m/\sqrt{2}$.   This agrees with equation (33) for  $\kappa =0$.
    The correspondence of the instability found here and the vortex
 sheet instability establishes that the instability found here is the supersonic
 Kelvin-Helmholtz instability.   Landau's (1944) study of the
 stability of a supersonic vortex sheet assumed $k_z=0$ and found linear stability for ${\cal M}_T > 2^{3/2}$.   However, under this
 condition linear instability occurs for $k_z \neq 0$ when
$  \left[k_\phi(k_\phi^2+k_z^2)^{-1/2}\right]{\cal M}_T  <  2^{3/2}$.

\begin{figure}[h]
    \centering
    \includegraphics[width=0.4\textwidth]{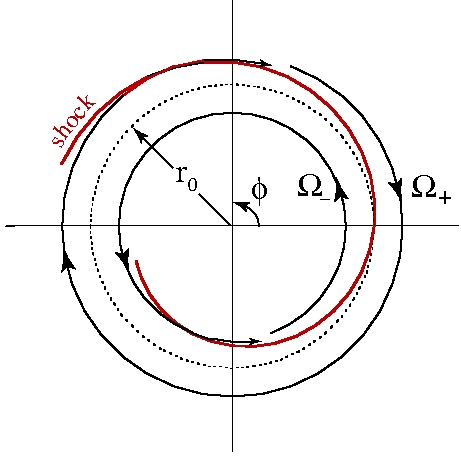}
    \caption{Polar plot of a one armed ($m=1$) perturbation which
 has steepened to form an oblique shock.  The gas crossing the shock
 is deflected towards the interface of the two components.}
\end{figure} 

\section{Nonlinear Evolution}

       Although the growth rate increases with increasing wavenumber
$k_z$, the large $k_z$ modes are not expected to be important because their
growth will saturate at a small amplitude $\delta r \sim k_z^{-1}$
and involve mixing only  small amounts of co- and counter-rotating matter.
For this reason we consider the long wavelength modes, $k_z \sim
h^{-1}$ and $k_\phi \geq r_0^{-1}$.   The  perturbations
have the form
\begin{equation}
\Psi_\pm \propto \exp\big[\mp \Re(k_{r\pm})(r-r_0)\big]
\exp\big[i(m\phi \mp \Im(k_{r\pm})(r-r_0)\big]~.
\end{equation}
The maximum of the perturbation is a tightly-wrapped
 leading spiral $r =r_0\pm m\phi/\Im(k_{r\pm})$ with $m$ arms.  
Figure 4 shows the nature of the spiral for $m=1$.

      The spiral wave can lead to the formation of an
oblique shock under some conditions.  Upstream of the
wave, the magnitude of the normal normal component of the gas velocity
relative to the wave is $u_n=\theta r_0 \Omega_0$, where $\theta = 
m| \Im(k_{r\pm})r_0|^{-1} \ll 1$.  
    If the upstream normal Mach number 
${\cal M}_n = u_n/c_s >1$ there is an oblique shock.  On the downstream side the normal Mach number is less than unity.  For
$r>r_0$ the gas is deflected inward and acquires a radial velocity
$u_r =-\beta r_0\Omega_0$, where $\beta =[2/(1+\gamma)](1-{\cal M}_n^{-2})\theta$.   For $r<r_0$, the oblique shock gives outward radial
velocity $u_r =\beta r_0\Omega_0$.

   The co- and counter-rotating disc components have a natural
{\it repulsion} for each other for the following reason.  
        The mixing of the two components
will involve strong shocks.   A possible geometry for $m=1$ is sketched in Figure 5 where there are two strong standing shocks, each with
Mach number ${\cal M} = u_\phi/c_s \gg 1$.  
            In the region
between the shocks, the angular momenta of the two components 
is annihilated and the gas is strongly heated.  The ratio
of the gas temperatures downstream to upstream of the normal shock is
$T_2/T_1 \approx 1+2(\gamma-1)\gamma {\cal M}^2/(\gamma+1)^2$.
This ratio is $\sim 20$ for  ${\cal M} =10$ and $\gamma=7/5$.  
The heated gas will expand vertically increasing its scale height
by a factor $\sqrt{T_2/T_1}\gg 1$.  With no centrifugal force,
the gas will free-fall over the surfaces of the inner disc towards the disc's center.

   The mass loss rate from this process is 
$\dt{M}_{\rm ann} = 4 m h r_0 \Delta r
 \rho \Omega_0$, where $\Delta r$ is the radial amplitude of the
 interface perturbation.   This loss rate is larger than the Shakura-Sunyaev
 (1973) accretion rate by a factor of order $(m/\pi)(\Delta r/h)({\cal M}/\alpha)$ where $\alpha$ is the dimensionless viscosity coefficient  for a 
 disc rotating in one direction.  It is commonly
 estimated to be in the range $10^{-3} - 0.1$.    The Kelvin-Helmholtz
 instability evidently gives a spatially localized effective viscosity coefficient $\alpha_{\rm eff} \sim (m/\pi) {\cal M}$ assuming $\Delta r \sim h$.
      Thus the region between the co- and counter-rotating components will
 be rapidly emptied of gas and a gap will form.  
     Subsequently, the outer disc will spread inward and the inner disc
outward due to the turbulent viscosity.  Consequently the gap will
close after a definite period and the mentioned heating and annihilation
of angular momentum will repeat.

\begin{figure}
    \centering
    \includegraphics[width=0.5\textwidth]{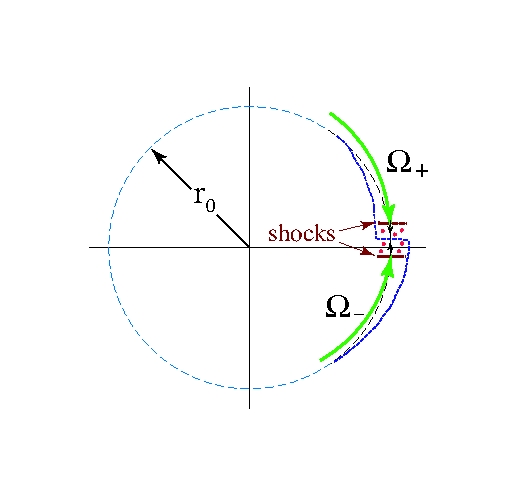}
    \caption{Polar plot suggesting the formation of strong normal
 shocks leading the annihilation of the angular momenta of the
 co- and counter rotating components and the heating of this gas.}
\end{figure} 

\section{Conclusions} 

    Observations of galaxies and models of accreting systems point to
the  occurrence of counter-rotating discs where the inner part of
the disc ($r<r_0$)  is co-rotating and the outer part is counter-rotating.
This work analyzes the linear stability of radially separated 
co- and counter-rotating thin discs.   The  strong instability is
the supersonic Kelvin-Helmholtz instability.  The growth rates are of the order of or larger than the angular rotation rate at the interface $\Omega(r_0)$.
  The instability is absent if there is no vertical dependence of the 
perturbation, that is, if $k_z=0$.   That is, the instability is essentially a
three-dimensional instability.   Thus this instability is not captured
by the analysis of Li and Narayan (2004) where $k_z=0$ and where the
gas is treated as incompressible.
Counter-rotating Keplerian discs
are unstable for azimuthal mode numbers $m=1,2,..$, whereas for
a flat rotation curve relevant to disc galaxies the unstable
mode numbers are $m=2,3,..$.
    We note that when there is a
 density change  across the interface there
is necessarily a corresponding change in the entropy.   This changes
the jump condition across the interface and leads in general to
propagating unstable modes where $\omega=\omega_R + i\omega_I$
with $\omega_R \neq 0$ and $\omega_I >0$.  The important new
term in the jump condition comes from the term $({\cal F}/\Omega
L_S^2)\Psi$ on the right-hand side of equation (17).

     The nonlinear phase of the instability will involve the mixing
 of co- and counter-rotating gas which will involve strong
 shocks, strong heating of the gas, and annihilation of the angular
 momenta of the two gas components.   The heated gas will 
 expand vertically and since it has zero  angular
 momentum it will free-fall to the disc's center over the
 surface of the inner disc. 
   Thus the region between the co- and counter-rotating components will
 be rapidly emptied of gas and a gap will form.  
   In effect the two components have a repulsive
 interaction.  This is suggestive of the Leidenfrost effect (Leidenfrost 1966)  where a liquid droplet is levitated above a hot surface by the
 vapor pressure between the droplet and the surface.
     Subsequent to the gap formation, the outer disc will spread inward and the inner disc outward due to the turbulent viscosity.  Consequently the gap will close after a definite period and the heating and annihilation of angular momentum will repeat.

    Three-dimensional hydrodynamic simulations are of course required
to fully understand the nonlinear evolution of the counter-rotating
discs.    Although axisymmetric disc simulations do not allow for the Kelvin-Helmholtz instability, the inclusion of viscosity can model the
effect of the turbulent viscosity due to the instability.   In an early study by Kuznetsov et al. (1999) the effective viscosity was the numerical viscosity due to the finite grid.  
In a recent study by Dyda et al. (2014), a high resolution grid was used with all components of the viscous stress tensor are included to model an isotropic Shakura-Sunyaev $\alpha-$viscosity.   Both studies clearly show the development of a gap between the co- and counter-rotating components and enhanced accretion.   The study by Dyda et al.
(2014) shows the quasi periodic closing and opening of the gap.

\section*{Acknowledgments}

    We thank M.M. Romanova for valuable discussions and
an anonymous referee for valuable comments.
     This work  was supported in part by NASA grants NNX11AF33G
and NSF grant AST-1211318.

 {}

\end{document}